\documentclass{article}

\usepackage{PRIMEarxiv}

\usepackage[utf8]{inputenc} % allow utf-8 input
\usepackage[T1]{fontenc}    % use 8-bit T1 fonts
\usepackage{hyperref}       % hyperlinks
\usepackage{url}            % simple URL typesetting
\usepackage{booktabs}       % professional-quality tables
\usepackage{amsfonts}       % blackboard math symbols
\usepackage{nicefrac}       % compact symbols for 1/2, etc.
\usepackage{microtype}      % microtypography
\usepackage{lipsum}
\usepackage{fancyhdr}       % header
\usepackage{graphicx}       % graphics
\graphicspath{{media/}}     % organize your images and other figures under media/ folder

\usepackage{amsmath}

\title{Impact of seed node position on network robustness under localized attacks
}

\author{
  Masaki Chujyo, Shu Liu, Fujio Toriumi\\
  Department of Systems Innovation, School of Engineering \\
  The University of Tokyo \\
  Tokyo Japan\\
  \texttt{mchujyo@g.ecc.u-tokyo.ac.jp} \\
}

\begin{document}
\maketitle

\begin{abstract}
Localized attacks (LAs), where damage propagates from a single seed node to its neighbors, pose significant threats to the robustness of complex networks. Although previous studies have extensively analyzed network vulnerability under such attacks, they typically assume random seed node placement and evaluate average robustness. However, the structural position of the seed node can significantly impact the extent of damage. This study proposes the Localized Attack Vulnerability Index (LAVI), a node-level metric that quantifies the potential impact of a LA initiated at a specific node. LAVI quantifies the cumulative number of severed links during attack progression, capturing how local connectivity and topological position amplify the resulting damage. Numerical experiments on synthetic and real-world networks demonstrate that LAVI correlates more strongly with network robustness degradation than standard centrality measures, such as degree, closeness, and betweenness. Our findings highlight that classical centrality metrics fail to capture key dynamics of spatially localized failures, while LAVI provides an accurate and generalizable indicator of node vulnerability under such disruptions.
\end{abstract}

% keywords can be removed
%\keywords{First keyword \and Second keyword \and More}

\section*{Introduction}

Modern societies increasingly depend on large-scale interconnected infrastructures, such as transportation and power grids, whose resilience to spatially concentrated disruptions from natural disasters poses a critical challenge \cite{boin2007preparing, mukherjee2014network, khademi2015transportation,tang2019addressing,kilanitis2019integrated,lam2021network}. Addressing this challenge requires models that systematically capture how localized damage can escalate into large-scale systemic failures. To model these dynamics, researchers have investigated localized attacks (LAs) on complex networks, where nodes are sequentially removed based on their topological or spatial proximity to a designated seed node \cite{shao2015percolation, berezin2015localized, vaknin2017spreading, dong2019localized}. These models are inspired by real-world phenomena, such as earthquakes, floods, and other spatially concentrated disruptions, where damage typically originates at a single point and spreads locally \cite{vaknin2017spreading, yan2017vulnerability, dong2022modest}.

Extensive research on network vulnerability under LAs has shown that even small initial disruptions can trigger abrupt fragmentation or cascading failures \cite{shao2015percolation, yuan2015breadth, berezin2015localized, vaknin2020spreading, lv2023cascading}. Recent studies have examined how network properties, such as degree distribution, clustering, community structure, and degree correlations, influence robustness to LAs, and have proposed mitigation strategies, including structural rewiring, localized recovery, and selective protection \cite{dong2019localized, geng2021global, fu2019node, shang2016localized, gong2019effective, chujyo2025improving}.

These theoretical insights are supported by empirical studies demonstrating that even modest localized disruptions can induce large-scale failures in spatial infrastructure systems, such as road networks and power grids \cite{hines2010topological, buldyrev2010catastrophic, zhou2019connectivity, wang2019local, dong2022modest}. For instance, empirical analyses of post-earthquake road networks have revealed that removing a few critical links in a localized region can severely degrade overall connectivity \cite{zhou2019connectivity}. Similarly, studies have demonstrated that localized flooding, despite its limited spatial scope, can trigger cascading failures in transportation networks owing to load redistribution and compound disruptions \cite{wang2019local, dong2022modest}. These findings highlight that the spatial concentration of initial damage, rather than its overall size, often determines systemic impact. Therefore, understanding the conditions under which localized damage escalates to network-wide failures is essential for designing resilient infrastructure systems.

Related efforts have also investigated node importance and dynamic processes in complex networks~\cite{artime2024robustness,schwarze2024structural}. In the context of targeted attacks, it is well-known that targeted removal of nodes in descending order of degree or betweenness centrality can rapidly fragment networks and significantly reduce global connectivity~\cite{albert2000error,holme2002attack}. Such strategies have therefore been widely used as benchmarks for assessing network robustness. Beyond classical centrality heuristics, a variety of optimization-based frameworks have been developed to identify structurally critical nodes for network dismantling and fragmentation, including optimal percolation~\cite{morone2015influence}, belief-propagation-based dismantling~\cite{mugisha2016identifying}, and machine-learning-driven network disintegration methods~\cite{grassia2021machine}. In parallel, in the context of cascading failures, several studies have reported strong correlations between node centrality measures and failure propagation severity and proposed centrality-based or hybrid strategies to identify critical nodes~\cite{ghanbari2018correlation,xiao2022cascading}. Moreover, related optimization and evolutionary approaches have also been explored for influence maximization and diffusion-oriented seed selection problems, aiming to identify influential and robust spreaders under information propagation models~\cite{wang2023enhancing,ou2024finding}.

These studies establish a broad framework for identifying structurally important nodes or influential spreaders under targeted removal and diffusion processes. The role of a seed in an LA, however, differs from that in these conventional settings.

In LAs, the seed specifies the origin of a localized removal process, and the damaged region expands outward according to topological or spatial proximity. Thus, the resulting damage depends not only on the structural importance of the seed itself but also on the shell structure surrounding it and on how connectivity is progressively removed as the affected region expands. This makes the identification of critical attack origins distinct from classical targeted attack and influence maximization problems.

Despite this potential dependence on the attack origin, most previous LA studies assume that the seed node is selected uniformly at random and evaluate robustness by averaging over multiple realizations~\cite{shao2015percolation,yuan2015breadth}. Such an approach is well suited for characterizing network-level robustness, but it averages out node-specific differences in the severity of localized attacks. This motivates us to examine vulnerability at the level of individual seed nodes and to ask whether the most damaging origins of localized disruption can be identified from network structure.

In this study, we address this gap and make three contributions. First, we clarify that vulnerability under localized attacks should be evaluated at the level of individual seed nodes, explicitly focusing on identifying critical origin nodes of localized disruption, rather than relying on network-level averages over randomly chosen seeds. Second, we introduce a seed-specific vulnerability metric tailored to the shell-wise removal process of localized attacks, providing a proxy for the damage associated with each possible attack origin. Third, through numerical experiments on both synthetic and real-world networks, we demonstrate that this perspective enables more accurate identification of vulnerable seed nodes than conventional centrality measures.

We introduce the Localized Attacks Vulnerability Index (LAVI), a metric designed to quantify the potential damage caused by a node when it serves as the seed of an LA. LAVI measures the cumulative extent of link removal induced during the attack process and serves as a proxy for the network's susceptibility to localized disruptions, enabling seed-level vulnerability assessment under localized attack dynamics.

\section*{Materials and methods}
To evaluate how a seed node’s position influences the severity of a LA, we conducted numerical experiments on both synthetic and real-world networks. This section describes the attack model, introduces the proposed vulnerability index, and details the methodological framework for evaluating network robustness under various conditions.

\subsection*{Localized Attack and Robustness Index}

To quantitatively evaluate network robustness under LA, we adopted the global robustness index \( R \), which measures the network’s ability to maintain connectivity during the node removal process~\cite{schneider2011mitigation,geng2021global}. For a given seed node \(i\) and node removal sequence \(\pi\), it is defined as the normalized cumulative size of the largest connected component (LCC):
\begin{equation}
R(i,\pi) = \frac{1}{N} \sum_{q=1}^{N} s_{\mathrm{LA}}(q \mid i,\pi),
\end{equation}
where \( N \) is the total number of nodes in the original network, and \( s_{\mathrm{LA}}(q \mid i,\pi) \) denotes the relative size of the LCC after the first \( q \) nodes in the removal sequence \(\pi\) have been removed.

The LA model simulates damage spreading from a local region within the network. For a given seed node \(i\), nodes are removed sequentially in layers based on their shortest-path distance \( l \) from the seed. Specifically, shell \( l = 0 \) contains the seed node itself, while \( l = 1 \) includes its immediate neighbors. Nodes are removed in random order within each shell, and the process continues outward until \( q \) nodes have been removed. Thus, once the seed \(i\) and removal sequence \(\pi\) are specified, the removed node set is uniquely determined at each step \(q\).

After each removal step, the LCC is computed, and its size normalized by \( N \) yields \( s_{\mathrm{LA}}(q \mid i,\pi) \), representing the fraction of nodes remaining in the LCC. When multiple random within-shell removal sequences are considered for the same seed node, the reported robustness is averaged over these realizations. A higher \( R \) value indicates greater robustness, reflecting the network’s ability to maintain LCCs throughout the attack.

It should be noted that our analysis focuses on structural robustness based on network connectivity and does not explicitly consider flow dynamics, capacity constraints, or domain-specific performance measures. Although real-world networks involve diverse operational mechanisms, maintaining basic connectivity between nodes is generally a necessary prerequisite for functional operation. For this reason, the size of the LCC has been widely used as a proxy for functional connectivity in network robustness studies~\cite{schneider2011mitigation,geng2021global}. In this study, we therefore evaluate network robustness under localized attacks by examining how the size of the LCC evolves during the attack process.

\subsection*{Localized Attack Vulnerability Index for a Seed Node}\label{sec:link-removal}

To quantify the influence of a seed node \(i\) on severity of a LA in \(G=(V,E)\), we introduce the LAVI \(\mathcal{L}(i)\). This metric is based on the cumulative number of half-links severed as the attack progresses through concentric distance shells, where each undirected edge contributes two half-links (one per endpoint).

First, we compute the shortest-path distances from the seed node \(i\) using a breadth-first search:
\[
  d_i(v)=\mathrm{dist}_G(i, v),\quad v\in V.
\]
Let $d_{\max}=\max_{v\in V} d_i(v)$ be the maximum distance. For each distance \(d\in \{0,1,\dots,d_{\max}\}\), we define the shell
\[
  S_d=\{\,v\in V: d_i(v)=d\,\}.
\]
Therefore, we construct the removal sequence
\[
  \pi(i)=\bigl[S_0,S_1,\dots,S_{d_{\max}}\bigr],
\]
by concatenating shells in non-decreasing order of distance from the seed node \(i\). Within each shell \(S_d\), nodes are processed based on two ordering schemes:

\begin{itemize}
  \item \textbf{Random-case removal}: nodes are randomly shuffled using a fixed random seed for reproducibility, yielding a uniformly random permutation.
  \item \textbf{Worst-case removal}: nodes are sorted in descending order of their original degrees \(\deg_G(v)\), such that high-degree nodes are deactivated first.
\end{itemize}
These two schemes represent different types of localized disruption scenarios. The random-case models stochastic local damage processes, such as those induced by natural hazards, where the internal failure order within an affected region is uncertain. In contrast, the worst-case removal approximates adversarial localized attacks that preferentially target structurally important nodes, providing an upper bound on potential damage severity.

\begin{figure}[tbp]
    %\begin{adjustwidth}{-2.25in}{0in}
    \centering
    \includegraphics[width=0.75\linewidth]{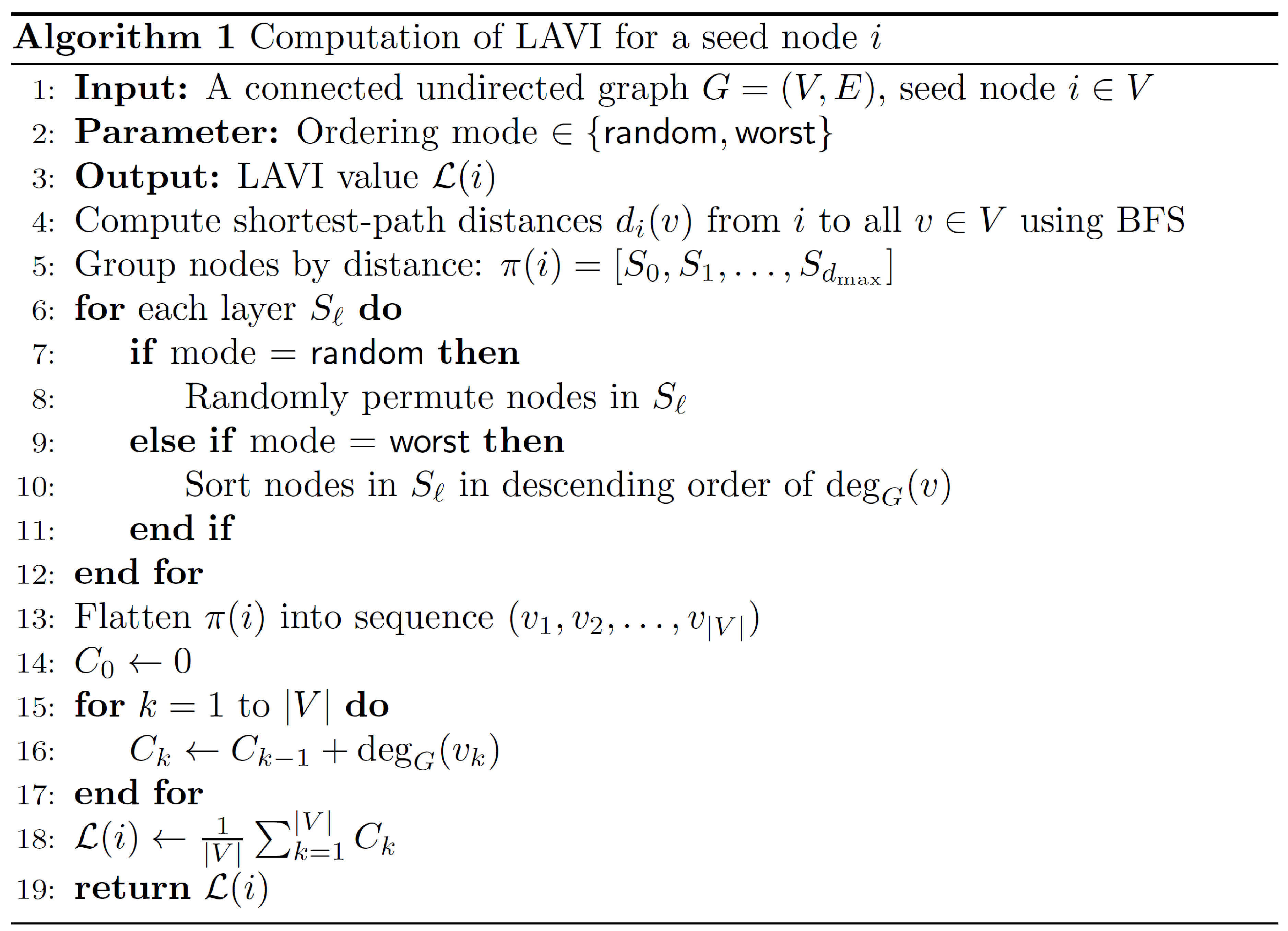}
    \caption{Pseudo-code for computing the LAVI value for a seed node $i$. Nodes are grouped by BFS distance from $i$, ordered within each layer, and evaluated via cumulative degree aggregation.}
    \label{fig:code}
    %\end{adjustwidth}
\end{figure}

Let \(\pi(i) = (v_1,v_2,\dots,v_{|V|})\) denote the full node removal sequence, ordered such that nodes closer to the seed \(i\) are removed earlier. After removing each node \(v_i\), we track the cumulative number of half-links severed up to step \(k\) as
\[
  C_k = \sum_{j=1}^k \deg_G(v_j), \quad k = 1,\dots,|V|.
\]
The LAVI \(\mathcal{L}(i)\) is defined as
\[
  \mathcal{L}(i) = \frac{1}{|V|} \sum_{k=1}^{|V|} C_k 
    = \frac{1}{|V|} \sum_{k=1}^{|V|} (|V|-k+1) \deg_G(v_k).
\]
The complete procedure for computing $\mathcal{L}(i)$ is described in pseudo-code form in Fig.~\ref{fig:code}.

%Intuitively, \(\mathcal{L}(i)\) is proportional to the area under the curve of cumulative half-link removals versus removal index; a larger \(\mathcal{L}(i)\) indicates that more links are removed earlier in the process, reflecting a severe attack initiated from the seed node \(i\). Note that dividing \(\mathcal{L}(i)\) by \(2M\) can yield a normalized value in the range \([0,1]\); however, we intentionally avoid this normalization, as it tends to compress the values excessively for sparse networks.

\begin{figure}[tbp]
  %\begin{adjustwidth}{-2.25in}{0in}
  \centering
  \includegraphics[width=0.7\linewidth]{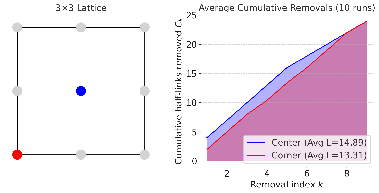}
  \caption{Random-case cumulative half-link removals \(C_k\) versus removal index \(k\) on a \(3\times3\) lattice. The blue curve shows the average over 10 runs when the seed node is placed at the center, while the red curve corresponds to placement at a corner. The shaded areas under each curve correspond to the LAVI \(\mathcal{L}(i)\). A larger shaded area (center seed) indicates more rapid link removal and hence greater vulnerability.}
  \label{fig:metric_example}
  %\end{adjustwidth}
\end{figure}

Intuitively, \(\mathcal{L}(i)\) is proportional to the area under the curve of cumulative half-link removals versus the removal index; a larger \(\mathcal{L}(i)\) indicates that more links are removed at earlier stages of the attack process, reflecting more severe damage initiated from the seed node \(i\). This design is motivated by the fact that network connectivity generally becomes more fragile as links are removed, and substantial early link loss under localized attacks tends to accelerate large-scale fragmentation. Unlike conventional centrality measures that capture static structural importance, \(\mathcal{L}(i)\) explicitly reflects the dynamic accumulation of connectivity loss during the outward propagation of localized failures. An illustrative $3\times3$ grid example is shown in Fig. \ref{fig:metric_example} to visualize the cumulative half-link removal process underlying \(\mathcal{L}(i)\).

%Note that dividing \(\mathcal{L}(i)\) by \(2M\) yields a normalized value in the range \([0,1]\); however, we intentionally avoid this normalization, as it tends to excessively compress the values in sparse networks and reduce discriminative resolution among seed nodes.

Note that dividing $\mathcal{L}(i)$ by $2M$ yields a normalized value in the range $[0,1]$. However, because $2M$ is constant for all candidate seed nodes within a given network, this normalization is simply a positive linear rescaling and therefore does not affect node rankings or the Pearson and Spearman correlations used in our analyses. Since the present study compares seed-node vulnerability within each network rather than absolute LAVI values across different networks, we report the unnormalized $\mathcal{L}(i)$.


The computation of \(\mathcal{L}(i)\) consists of (i) breadth-first search for distance calculation in \(O(|V|+|E|)\) time; (ii) shell construction in \(O(|V|)\); (iii) within-shell ordering, requiring \(O(|V|\log|V|)\) time in the worst-case removal (assuming worst-case shell size), or \(O(|V|)\) time in the random-case removal via random shuffling; and (iv) cumulative sum calculation in \(O(|V|)\). The total time complexity is \(O(|V|\log|V|+|E|)\) for worst-case removal and \(O(|V|+|E|)\) for random-case removal.

\section*{Results}
To evaluate how the structural position of a seed node affects the outcome of LAs, we performed simulations on both synthetic and real-world networks. Synthetic networks enable controlled variation of structural properties, while real-world networks demonstrate the applicability of our findings to practical systems. The results are presented in two parts: first for synthetic networks, then for real-world datasets.

\subsection*{Results for Synthetic Networks}

To investigate the relationship between network structure and vulnerability to LAs, we performed a systematic analysis using the Barab\'asi–Albert (BA) model~\cite{barabasi1999emergence}, a widely used generative model for scale-free networks. Specifically, we generated BA networks with \( N = 100 \) nodes and attachment parameter \( m = 2 \), yielding sparse yet connected topologies with heterogeneous degree distribution.

We analyzed the impact of LAs initiated from different seed nodes by randomly sampling 100 seed nodes from the network. For each sampled seed, we conducted 100 independent simulations of LA and computed the average robustness index \( R \). This procedure yields a stable quantification of the structural damage associated with attacks initiated from various topological positions.

To understand which node-level properties are most indicative of vulnerability, we computed two versions of the proposed LAVI: \( \mathcal{L}_{\text{worst}} \) and \( \mathcal{L}_{\text{random}} \), along with classical centrality measures, including degree, closeness, and betweenness centrality \cite{freeman2002centrality}.

\begin{figure}[thpb]
    %\begin{adjustwidth}{-2.25in}{0in}
    \centering
    \includegraphics[width=0.99\linewidth]{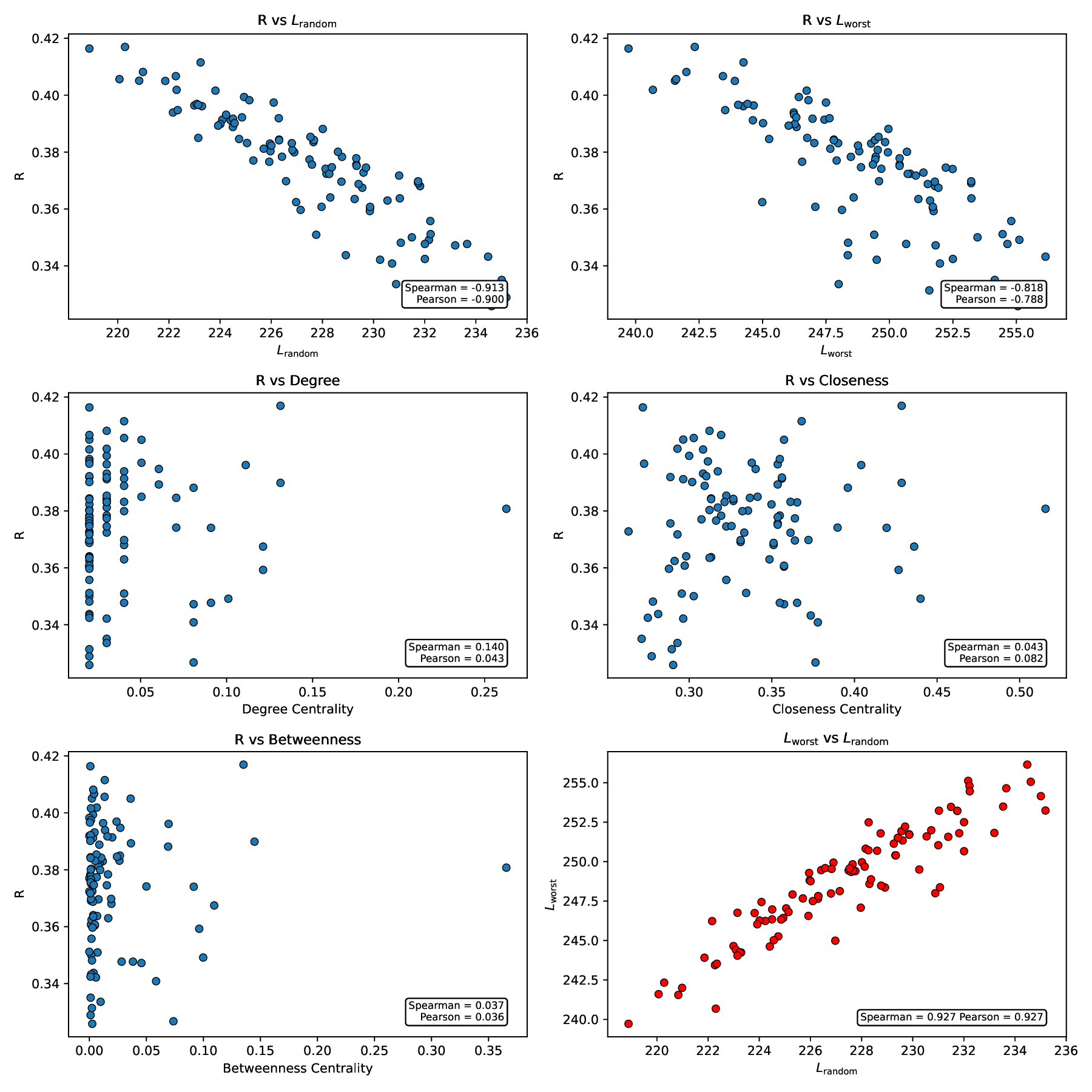}
    \caption{Correlation between the robustness index $R$ and various indices characterizing the choice of the seed node in a Barab\'asi--Albert network with $N = 100$ and $m = 2$. The panels show correlations between $R$ and $\mathcal{L}_{\text{random}}$, $\mathcal{L}_{\text{worst}}$, degree, closeness, and betweenness centrality of the seed node, as well as the mutual correlation between $\mathcal{L}_{\text{worst}}$ and $\mathcal{L}_{\text{random}}$. Spearman and Pearson correlation coefficients are reported in each panel.}
    \label{fig:results_BA_N100_m2}
    %\end{adjustwidth}
\end{figure}

Figure~\ref{fig:results_BA_N100_m2} presents the correlations between the robustness index \( R \) and various node-level metrics. The top row displays results for the two vulnerability indices and their mutual correlation, while the bottom row displays correlations with the three centrality measures.

The top row results show that both vulnerability indices are strongly negatively correlated with robustness. Higher values of \( \mathcal{L}_{\text{random}} \) and \( \mathcal{L}_{\text{worst}} \) are associated with lower values of \( R \), indicating that seed nodes causing faster cumulative link removals tend to cause greater structural damage. Among them, \( \mathcal{L}_{\text{random}} \) shows the strongest correlation (Spearman \( \rho = -0.913 \), Pearson \( r = -0.900 \)), followed by \( \mathcal{L}_{\text{worst}} \) (Spearman \( \rho = -0.818 \), Pearson \( r = -0.788 \)). The high agreement between these two indices (Spearman and Pearson \( \rho = 0.927 \)) indicates that the structural position of the seed node plays a dominant role, irrespective of within-shell removal order.

In contrast, the classical centrality measures (bottom row) show weak or negligible correlations with robustness. Degree centrality exhibits virtually uncorrelated with \( R \), while closeness and betweenness fail to explain robustness variations. These findings suggest that traditional centrality metrics are ineffective in identifying structurally vulnerable nodes under LA scenarios.

Overall, these results highlight the effectiveness of the proposed LAVI \( \mathcal{L}(i) \) in identifying seed nodes cause the greatest reduction in robustness. By capturing both the spatial progression of the attack and the rate of link removal, the index provides a more relevant assessment of vulnerability than conventional centrality measures.

To assess the scalability of our findings, we extended the analysis to BA networks with \( N = 1000 \) and \( N = 10000 \), using the same attachment parameter \( m = 2 \). Table~\ref{tab:BA_correlation_summary} summarizes the Spearman and Pearson correlations between \( R \) and the various node-level indices for each network size.

\begin{table}[htbp]
%\begin{adjustwidth}{-2.25in}{0in}
\centering
\caption{Spearman and Pearson correlations between robustness index \( R \) and node-level indices in BA networks of varying sizes (\( N = 100, 1000, 10000 \), \( m = 2 \)). Bold indicates the strongest (most negative) correlation for each network.}
\label{tab:BA_correlation_summary}
\begin{tabular}{llrr}
\hline
\textbf{Network} & \textbf{Index} & \textbf{Spearman} & \textbf{Pearson} \\
\hline
$N=100$ 
  & LAVI(random)  & \(-\mathbf{0.913}\) & \(-\mathbf{0.900}\) \\
  & LAVI(worst)   & \(-\mathbf{0.818}\) & \(-\mathbf{0.788}\) \\
  & Degree        & \(0.140\)           & \(0.043\) \\
  & Closeness     & \(0.043\)           & \(0.082\) \\
  & Betweenness   & \(0.037\)           & \(0.036\) \\
\hline
$N=1000$
  & LAVI(random)  & \(-\mathbf{0.693}\) & \(-\mathbf{0.717}\) \\
  & LAVI(worst)   & \(-\mathbf{0.663}\) & \(-\mathbf{0.692}\) \\
  & Degree        & \(-0.031\)          & \(-0.033\) \\
  & Closeness     & \(-0.009\)          & \(-0.083\) \\
  & Betweenness   & \(-0.081\)          & \(-0.062\) \\
\hline
$N=10000$ 
  & LAVI(random)  & \(-\mathbf{0.570}\) & \(-\mathbf{0.590}\) \\
  & LAVI(worst)   & \(-\mathbf{0.464}\) & \(-\mathbf{0.488}\) \\
  & Degree        & \(-0.033\)          & \(0.013\) \\
  & Closeness     & \(0.181\)           & \(0.138\) \\
  & Betweenness   & \(0.062\)           & \(0.023\) \\
\hline
\end{tabular}
%\end{adjustwidth}
\end{table}

Across all network sizes, both versions of the vulnerability index, especially \( \mathcal{L}_{\text{random}} \), maintain strong negative correlations with robustness. However, as the network size increases, these correlations gradually weaken—for example, the Spearman correlation of \( \mathcal{L}_{\text{random}} \) with \( R \) drops from \(-0.91\) at \( N = 100 \) to \(-0.57\) at \( N = 10000 \). Therefore, while LAVI remains a reliable predictor of vulnerability, its ability to differentiate may decline in larger, more heterogeneous networks, where local effects weaken.

In contrast, classical centrality measures consistently yield weak or negligible correlations across all scales. Therefore, these results further elaborate that the proposed vulnerability indices outperform traditional measures in identifying structurally critical nodes under LAs. This phenomenon holds across different network sizes.

To examine whether the correlation between LAVI and robustness generalizes beyond BA networks, we extended the analysis to these five synthetic network models. The synthetic networks considered here include Barab\'asi--Albert (BA) networks~\cite{barabasi1999emergence}, Erd\H{o}s--R\'enyi (ER) networks~\cite{Erdos1984OnTE}, Watts--Strogatz (WS) networks~\cite{watts1998collective}, stochastic block model (SBM) networks~\cite{holland1983stochastic}, and random geometric graphs (RGGs)~\cite{penrose2003random}, representing heterogeneous, homogeneous random, small-world, modular, and spatially embedded network structures, respectively. To facilitate a controlled comparison of the effects of network topology, we fixed the network size at $N=1000$ and selected the model parameters such that the mean degree was approximately $\langle k\rangle=6$ across the different network models.

Table~\ref{tab:synthetic_top10_corr} summarizes the Spearman rank correlations between robustness and each node-level metric over 10 network realizations. LAVI exhibits stronger negative correlations than the conventional centrality measures in the BA, ER, WS, and SBM networks, although the relative performance of the two LAVI variants varies across network models. In contrast, closeness centrality shows the strongest negative correlation in the RGG network, indicating that the predictive performance of LAVI depends on network topology.

\begin{table}[t]
%\begin{adjustwidth}{-2.25in}{0in}
\centering
\caption{Spearman rank correlations between robustness under localized attacks and each vulnerability metric for synthetic networks. }
\label{tab:synthetic_top10_corr}
\begin{tabular}{llrrrrr}
\hline
Network & Parameter & $\mathcal{L}_\mathrm{worst}$ & $\mathcal{L}_\mathrm{random}$ & Degree & Closeness & Betweenness \\
\hline
BA  & $m=3$                       & $-0.532$ & $\textbf{-0.686}$ & $-0.106$ & $-0.270$ & $-0.161$ \\
ER  & $\langle k\rangle\approx6$ & $\textbf{-0.544}$ & $-0.518$ & $-0.035$ & $-0.084$ & $-0.016$ \\
WS  & $p=0.1$                     & $\textbf{-0.322}$ & $-0.192$ & $-0.030$ & $-0.102$ & $-0.018$ \\
SBM & $\mu=0.15$                  & $\textbf{-0.623}$ & $-0.482$ & $0.086$  & $-0.046$ & $-0.025$ \\
RGG & $\langle k\rangle\approx6$ & $-0.252$ & $-0.208$ & $-0.042$ & $\textbf{-0.607}$ & $-0.182$ \\
\hline
\end{tabular}
%\end{adjustwidth}
\end{table}

\subsubsection*{Identification of Critical Seeds on Synthetic Networks}

Correlation analysis quantifies the overall association between a node-level metric and attack severity, but does not directly evaluate whether the metric can identify the most damaging seed nodes. To assess this practical aspect, we additionally ranked all nodes in descending order according to each vulnerability metric and centrality measure and selected the top 10\% of nodes as candidate critical seeds. Localized attacks were then initiated from these selected seeds, and the resulting mean robustness was compared across the different ranking methods. Because a smaller value of $R$ corresponds to greater structural damage, a metric that successfully identifies vulnerable seeds is expected to yield a lower mean $R$.

Table~\ref{tab:synthetic_top10_topk} summarizes the results for the five synthetic network models. In the BA network, $\mathcal{L}_{\mathrm{random}}$ produced the smallest mean robustness, $R=0.4155$, yielding lower mean robustness than degree, closeness, betweenness, and random seed selection. In the ER, WS, and SBM networks, both LAVI variants also yielded lower mean robustness than the conventional centrality measures. These results indicate that the ranking performance of LAVI is not restricted to scale-free BA networks and extends to networks with homogeneous degree distributions, small-world structure, and community structure.

A different tendency was observed for the RGG network. In this spatially embedded topology, closeness centrality yielded the smallest robustness, $R=0.4063$, whereas the LAVI-based rankings resulted in $R=0.4275$--$0.4332$. This result suggests that the global structural position of a seed can become particularly important in spatially embedded networks. Thus, although LAVI effectively identifies damaging seeds across several network classes, its predictive advantage is topology dependent and may decrease when the global position of the seed plays a dominant role in fragmentation.

\begin{table}[t]
%\begin{adjustwidth}{-2.25in}{0in}
\centering
\caption{Mean robustness $R$ obtained from localized attacks initiated from the top 10\% of nodes ranked by each node-level metric in synthetic networks. Lower values indicate more damaging seed selections.}
\label{tab:synthetic_top10_topk}
\begin{tabular}{llrrrrrr}
\hline
Network & Parameter & $\mathcal{L}_{\mathrm{worst}}$ & $\mathcal{L}_{\mathrm{random}}$ & Degree & Closeness & Betweenness & Random \\
\hline
BA  & $m=3$                       & $0.4180$ & $\textbf{0.4155}$ & $0.4241$ & $0.4252$ & $0.4245$ & $0.4218$ \\
ER  & $\langle k\rangle\approx6$ & $\textbf{0.4541}$ & $0.4542$ & $0.4565$ & $0.4562$ & $0.4567$ & $0.4568$ \\
WS  & $p=0.1$                     & $\textbf{0.4607}$ & $0.4617$ & $0.4644$ & $0.4636$ & $0.4647$ & $0.4641$ \\
SBM & $\mu=0.15$                  & $\textbf{0.4600}$ & $0.4606$ & $0.4640$ & $0.4630$ & $0.4636$ & $0.4639$ \\
RGG & $\langle k\rangle\approx6$ & $0.4275$ & $0.4332$ & $0.4509$ & $\textbf{0.4063}$ & $0.4383$ & $0.4524$ \\
\hline
\end{tabular}
%\end{adjustwidth}
\end{table}

\subsection*{Results for Real-World Networks}

To assess the practical relevance of our findings, we performed LA analysis to seven real-world networks drawn from diverse domains, including infrastructure, biological systems, social networks, and cheminformatics. The datasets, listed in Table~\ref{tab:dataset}, vary widely in size, density, and assortativity. This enables easy examination of the generality of the proposed vulnerability indices under heterogeneous conditions.

\begin{table}[bp]
%\begin{adjustwidth}{-2.25in}{0in}
\centering
\caption{Summary statistics of the real-world networks used in the analysis.}
\label{tab:dataset}
\begin{tabular}{lrrrrr}
\hline
\textbf{Network Name} & \textbf{Type} & \textbf{\#Nodes} & \textbf{\#Edges} & \textbf{Assortativity} & \textbf{Reference} \\
\hline
\texttt{NCI1\_g4094}          & Cheminformatics      & 90    & 98     & -0.286 & \cite{NDR} \\
\texttt{inf-USAir97}          & Infrastructure       & 332   & 2126   & -0.208 & \cite{NDR,colizza2007reaction} \\
\texttt{power-494-bus}        & Power grid           & 494   & 586    & -0.074 & \cite{NDR} \\
\texttt{bio-diseasome}        & Biological           & 516   & 1188   & 0.067  & \cite{NDR,goh2007human} \\
\texttt{fb-pages-food}        & Social               & 620   & 2102   & -0.028 & \cite{NDR,rozemberczki2019gemsec} \\
\texttt{bio-yeast}            & Biological           & 1458  & 1948   & -0.210 & \cite{NDR,jeong2001lethality} \\
\texttt{inf-openflights}      & Infrastructure       & 2939  & 15677  & 0.051  & \cite{NDR,opsahl2010node} \\
\hline
\end{tabular}
%\end{adjustwidth}
\end{table}

\begin{figure}[tbp]
    %\begin{adjustwidth}{-2.25in}{0in}
    \centering
    \includegraphics[width=0.99\linewidth]{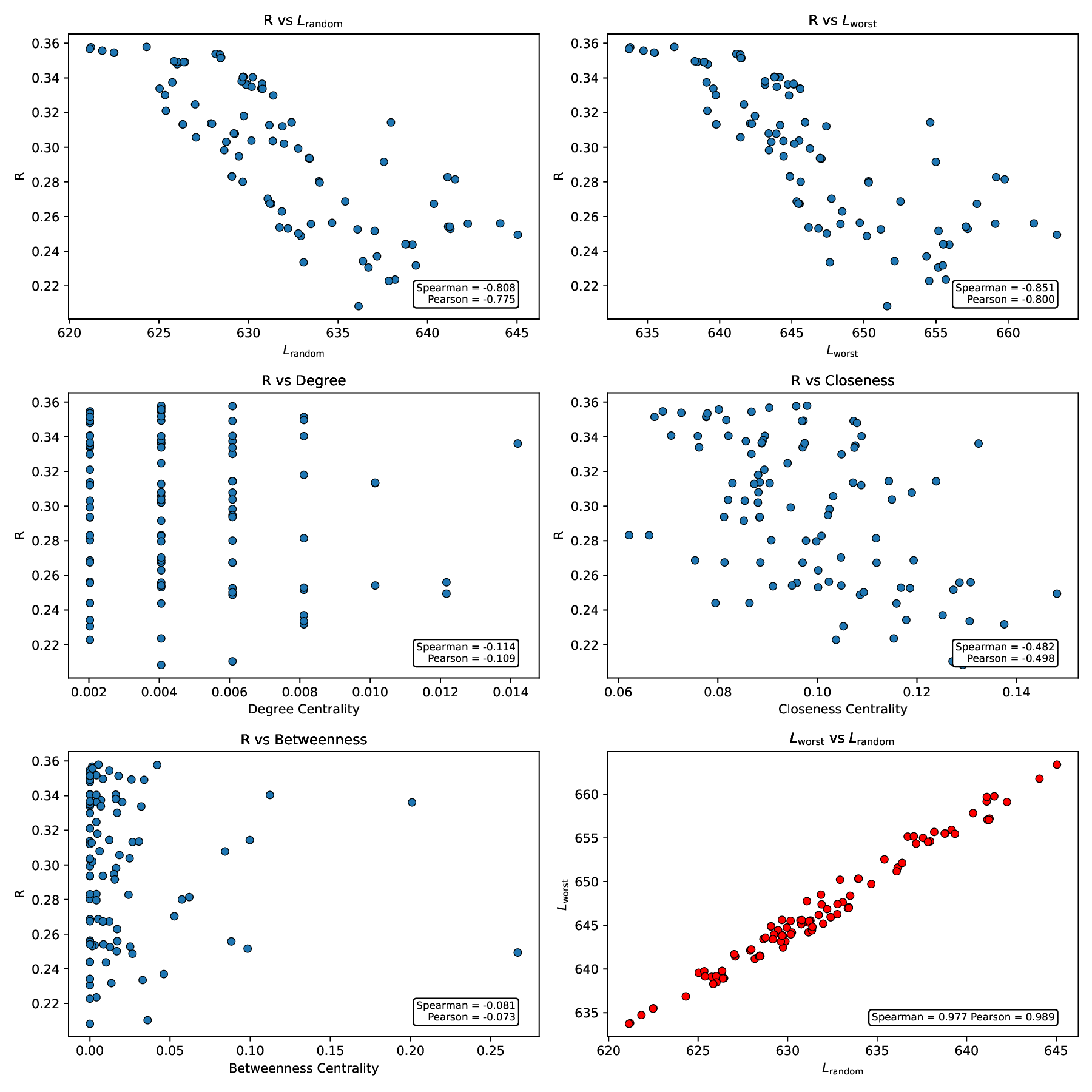}
    \caption{Correlation between the robustness index $R$ and various indices characterizing the choice of the seed node in the \texttt{power-494-bus} network. The panels demonstrate correlations between $R$ and $\mathcal{L}_{\text{random}}$, $\mathcal{L}_{\text{worst}}$, degree, closeness, and betweenness centrality of the seed node, as well as the mutual correlation between $\mathcal{L}_{\text{worst}}$ and $\mathcal{L}_{\text{random}}$. Spearman and Pearson correlation coefficients are reported in each panel.}
    \label{fig:power-494-bus}
    %\end{adjustwidth}
\end{figure}

Figure~\ref{fig:power-494-bus} presents results for one representative case, the \texttt{power-494-bus} network. The top row confirms that both \( \mathcal{L}_{\text{worst}} \) and \( \mathcal{L}_{\text{random}} \) are strongly negatively correlated with the robustness index \( R \), which is consistent with our findings in BA synthetic networks. The bottom row indicates that traditional centrality measures, degree, closeness, and betweenness, exhibit negligible or weak associations with robustness.

\begin{table}[t]
%\begin{adjustwidth}{-2.25in}{0in}
\centering
\caption{Spearman correlations between the robustness index $R$ and seed node indices. Values in parentheses are p-values. Bold shows the strongest correlation.}
\label{Spearman_real}
\begin{tabular}{lccccc}
\hline
\textbf{Network} & \textbf{LAVI(worst)} & \textbf{LAVI(random)} & \textbf{Degree} & \textbf{Closeness} & \textbf{Betweenness} \\
\hline
\texttt{NCI1\_g4094}       & \textbf{-0.90 (5.9e-33)} & -0.60 (4.0e-10) &  0.03 (7.6e-01) & -0.70 (1.4e-14) & -0.16 (1.3e-01) \\
\texttt{inf-USAir97}       & -0.42 (1.3e-05)          & \textbf{-0.58 (1.9e-10)} &  0.02 (8.8e-01) & -0.22 (2.6e-02) & -0.09 (3.6e-01) \\
\texttt{power-494-bus}     & \textbf{-0.85 (3.2e-29)} & -0.81 (2.8e-24) & -0.11 (2.6e-01) & -0.48 (3.8e-07) & -0.08 (4.2e-01) \\
\texttt{bio-diseasome}     & -0.10 (3.1e-01)          &  0.20 (5.0e-02) & -0.01 (9.3e-01) & \textbf{-0.31 (2.0e-03)} & -0.26 (1.0e-02) \\
\texttt{fb-pages-food}     & \textbf{-0.56 (1.3e-09)} & -0.50 (9.4e-08) &  0.15 (1.5e-01) & -0.20 (5.1e-02) &  0.00 (9.9e-01) \\
\texttt{bio-yeast}         & \textbf{-0.75 (3.2e-19)} & -0.63 (2.7e-12) &  0.03 (7.7e-01) & -0.16 (1.1e-01) &  0.07 (5.1e-01) \\
\texttt{inf-openflights}   & \textbf{-0.53 (1.5e-08)} & -0.42 (1.2e-05) &  0.20 (4.3e-02) & -0.01 (9.3e-01) &  0.20 (4.4e-02) \\
\hline
\end{tabular}
%\end{adjustwidth}
\end{table}

\begin{table}[thb]
%\begin{adjustwidth}{-2.25in}{0in}
\centering
\caption{Pearson correlations between the robustness index $R$ and seed node indices. Values in parentheses are p-values. Bold shows the strongest correlation.}
\label{Pearson_real}
\begin{tabular}{lccccc}
\hline
\textbf{Network} & \textbf{LAVI(worst)} & \textbf{LAVI(random)} & \textbf{Degree} & \textbf{Closeness} & \textbf{Betweenness} \\
\hline
\texttt{NCI1\_g4094}       & \textbf{-0.92 (4.7e-37)} & -0.73 (4.1e-16) &  0.08 (4.6e-01) & -0.75 (1.9e-17) & -0.27 (9.7e-03) \\
\texttt{inf-USAir97}       & -0.59 (8.4e-11)          & \textbf{-0.64 (6.4e-13)} & -0.13 (1.8e-01) & -0.37 (1.5e-04) & -0.18 (8.1e-02) \\
\texttt{power-494-bus}     & \textbf{-0.80 (1.8e-23)} & -0.77 (3.1e-21) & -0.11 (2.8e-01) & -0.50 (1.3e-07) & -0.07 (4.7e-01) \\
\texttt{bio-diseasome}     & -0.11 (2.9e-01)          &  0.11 (2.6e-01) &  0.05 (6.4e-01) & \textbf{-0.37 (1.6e-04)} & -0.18 (7.8e-02) \\
\texttt{fb-pages-food}     & \textbf{-0.66 (6.8e-14)} & -0.46 (1.6e-06) &  0.05 (6.5e-01) & -0.16 (1.2e-01) & -0.05 (6.3e-01) \\
\texttt{bio-yeast}         & \textbf{-0.77 (1.6e-20)} & -0.67 (2.2e-14) & -0.01 (9.3e-01) & -0.21 (3.7e-02) & -0.02 (8.8e-01) \\
\texttt{inf-openflights}   & \textbf{-0.98 (1.0e-73)} & -0.97 (1.8e-62) & -0.02 (8.3e-01) & -0.48 (3.9e-07) & -0.02 (8.5e-01) \\
\hline
\end{tabular}
%\end{adjustwidth}
\end{table}

To systematically compare across all real-world networks, Tables~\ref{Spearman_real} and~\ref{Pearson_real} report Spearman and Pearson correlations between the robustness index \( R \) and the node-level indices. In most cases, \( \mathcal{L}_{\text{worst}} \) exhibits the strongest (most negative) correlation with robustness, highlighting its effectiveness as a predictor of vulnerability under LA. \( \mathcal{L}_{\text{random}} \) performs well, often outperforming traditional centrality measures despite not being optimized for worst-case shell removal.

\begin{figure}[th]
    %\begin{adjustwidth}{-2.25in}{0in}
    \centering
    \includegraphics[width=0.99\linewidth]{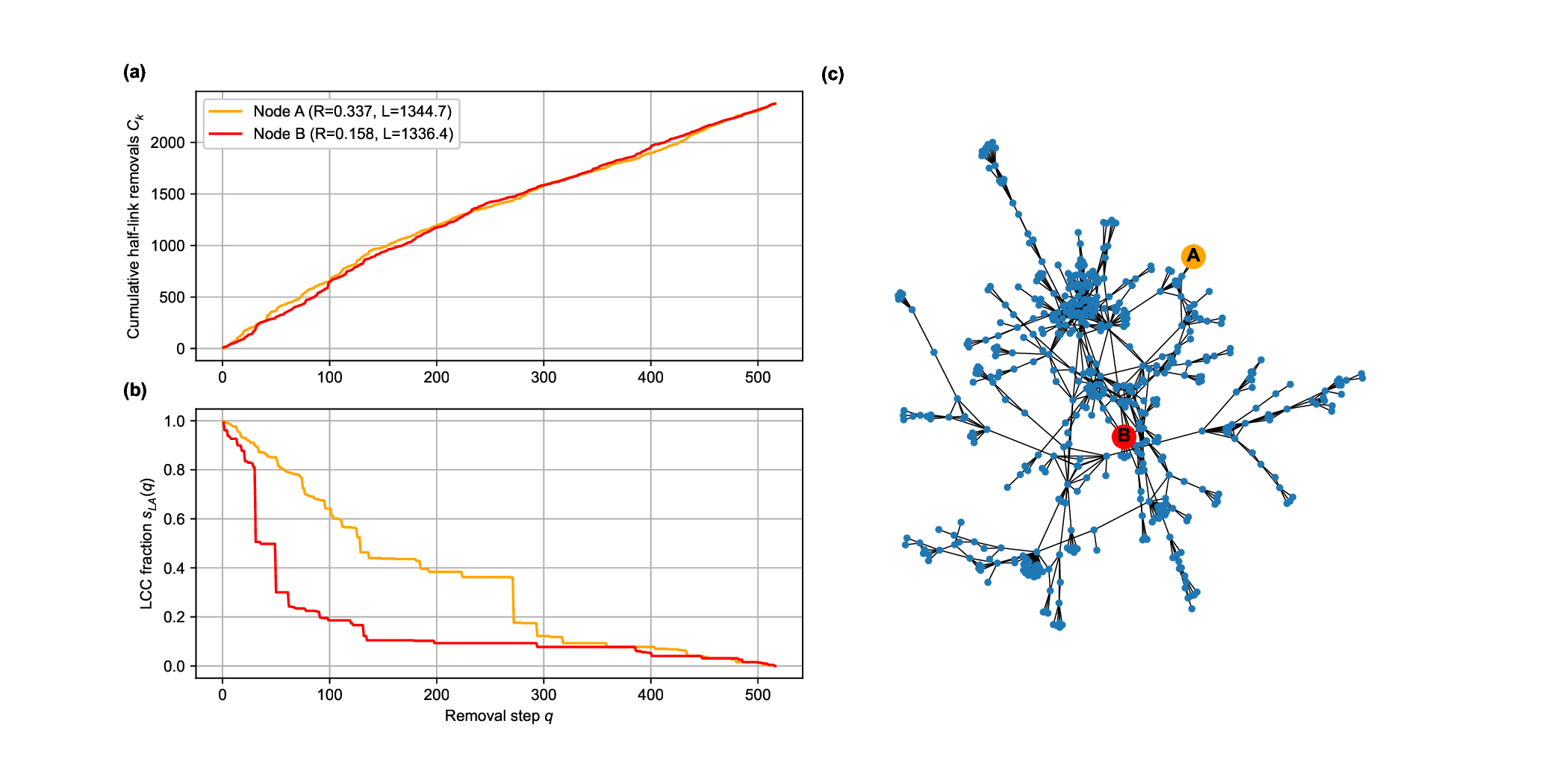}
    \caption{ Example from the bio-diseasome network illustrating a case where similar LAVI values correspond to substantially different robustness outcomes under localized attack. 
    (a) Cumulative half-link removals $C_k$ and (b) largest connected component fraction $s_\mathrm{LA}$ for two seed nodes with similar LAVI values. (c) Network visualization highlighting the positions of the two seeds.}
    \label{fig:limited_lavi}
    %\end{adjustwidth}
\end{figure}

This overall trend is consistent with the synthetic-network results, indicating that the proposed indices retain predictive ability across a range of real-world network structures.
However, certain exceptions emerge, as in networks such as bio-diseasome, where LAVI exhibits only a weak correlation with robustness (approximately $-0.1$). To further investigate this behavior, Fig.~\ref{fig:limited_lavi} presents a representative example from the bio-diseasome network, in which two seed nodes, A and B, exhibit nearly identical $\mathcal{L}_{\text{random}}=1344$ and $1336$, while their robustness indices under localized attacks differ substantially ($R_\mathrm{LA}=0.34$ and $0.16$). Although the cumulative half-link removals remain comparable (Fig.~\ref{fig:limited_lavi}(a)), the evolution of the largest connected component shows pronounced differences (Fig.~\ref{fig:limited_lavi}(b)), indicating that similar local vulnerability scores do not necessarily translate into equivalent global fragmentation outcomes.

The network visualization in Fig.~\ref{fig:limited_lavi}(c) further reveals that node B is located near the network core, whereas node A lies closer to the periphery, and that the network is composed of multiple small clusters bridged by a limited number of inter-community connections. In such topologies, the disruption of bridging links, rather than the total number of removed links, becomes the dominant factor governing large-scale fragmentation. This mechanism is consistent with previous observations suggesting that attacks targeting high-betweenness nodes can, in some cases, outperform degree-based strategies in real-world networks~\cite{holme2002attack}. In networks with modular structure and bridge-dependent connectivity, LAVI may exhibit weaker correlations with robustness, reflecting a structural limitation of the proposed approach.

These results also help explain why, among conventional centrality measures, closeness centrality exhibits relatively better performance in the bio-diseasome network. As shown in Fig.~\ref{fig:limited_lavi}(c), this network is characterized by modular structure and bridge-like connections, where damage propagation strongly depends on the relative position of the seed within the global topology. In such settings, seeds located closer to the network core can more easily affect multiple communities, leading to more severe fragmentation.

In contrast, degree and betweenness centralities primarily capture local connectivity patterns or shortest-path structure, which are effective in classical targeted attack scenarios but less informative for predicting the global impact of expanding localized failures. As a result, conventional centrality measures exhibit limited correlation with robustness under localized attacks. These observations highlight the necessity of propagation-aware metrics such as LAVI, which explicitly account for the cumulative and evolving nature of localized attack processes.

These observations indicate that the overall correlation between LAVI and robustness may become weak for certain network structures; however, this does not necessarily imply that LAVI cannot identify particularly vulnerable seed nodes.

\subsubsection*{Identification of Critical Seeds in Real-World Networks}

To examine whether the critical-seed identification performance observed in the synthetic networks also holds for empirical network structures, we applied the same top-10\% evaluation to the seven real-world networks. For each network, nodes were ranked according to each vulnerability metric and centrality measure, and the mean robustness was evaluated for localized attacks initiated from the top-ranked nodes.

Table~\ref{tab:real_top10} summarizes the results. Consistent with the synthetic-network results, the LAVI-based rankings yielded the lowest or tied-lowest mean robustness in six of the seven networks. Notably, in the bio-diseasome network, where LAVI exhibited only a weak overall correlation with robustness, $\mathcal{L}_{\mathrm{worst}}$ nevertheless yielded the lowest mean robustness in the top-10\% evaluation. This result suggests that a weak global correlation does not necessarily imply poor performance in identifying the most critical subset of seed nodes. An exception was the power-494-bus network, where closeness centrality yielded the lowest mean robustness, further indicating that the relative performance of LAVI depends on network topology.

\begin{table}[ht]
%\begin{adjustwidth}{-2.25in}{0in}
\centering
\caption{Mean robustness $R$ obtained from localized attacks initiated from the top 10\% of nodes ranked by each node-level metric in real-world networks. Lower values indicate more damaging seed selections.}
\label{tab:real_top10}
\begin{tabular}{lcccccc}
\hline
Network & $\mathcal{L}_{\mathrm{worst}}$ & $\mathcal{L}_{\mathrm{random}}$ & Degree & Closeness & Betweenness & Random \\
\hline
NCI1\_g4094      & \textbf{0.3646} & \textbf{0.3646} & 0.4035 & 0.4054 & 0.4196 & 0.4050 \\
inf-USAir97      & \textbf{0.2591} & 0.2651 & 0.2712 & 0.2681 & 0.2794 & 0.2886 \\
power-494-bus    & 0.2660 & 0.2615 & 0.2956 & \textbf{0.2581} & 0.2912 & 0.3018 \\
bio-diseasome    & \textbf{0.2066} & 0.2422 & 0.2637 & 0.2120 & 0.2376 & 0.2485 \\
fb-pages-food    & \textbf{0.3016} & 0.3093 & 0.3337 & 0.3307 & 0.3317 & 0.3323 \\
bio-yeast        & \textbf{0.2278} & 0.2311 & 0.2505 & 0.2475 & 0.2528 & 0.2532 \\
inf-openflights  & 0.2380 & \textbf{0.2372} & 0.2460 & 0.2405 & 0.2441 & 0.2571 \\
\hline
\end{tabular}
%\end{adjustwidth}
\end{table}

\section*{Discussion}

Our analysis reveals that the structural vulnerability of complex networks to LAs is not well captured by conventional centrality measures of seed nodes. Although high centrality is often presumed to indicate vulnerability, degree and betweenness centralities exhibit weak associations with robustness loss under LAs, while closeness centrality shows only moderate and inconsistent performance. These results indicate that node importance under LAs cannot be explained solely by global prominence or shortest-path accessibility.

Instead, the severity of a LA depends on how the removal of a seed node initiates and sustains the breakdown of nearby connectivity during outward propagation. LAVI captures this mechanism by quantifying the cumulative loss of links as the attack expands in a shell-wise manner, thereby incorporating both the topological progression of damage and the structural exposure of the network. This propagation-aware perspective explains why LAVI generally outperforms conventional centrality measures across the synthetic and real-world networks considered in this study.

At the same time, the predictive performance of LAVI varies across networks with different topological structures. In scale-free networks such as those generated by the BA model, LAVI is particularly effective because early link removal directly contributes to global connectivity loss, as high-degree nodes play a dominant role in maintaining connectivity. The reduction in predictive performance observed for larger networks may be related not simply to network size itself, but also to the characteristic shell structure of small-world networks. Since typical shortest-path distances increase only slowly with network size, a localized attack can reach a substantial fraction of the network within a relatively small number of shells. Because LAVI aggregates cumulative link loss over the entire removal sequence, seed-specific differences originating from the early local shells may become relatively diluted once the attack reaches larger shells containing substantial portions of the network. A similar tendency is observed in many real-world networks, which often exhibit heterogeneous or heavy-tailed degree distributions, enabling LAVI to effectively capture vulnerability patterns. The additional SBM results also indicate that the presence of community structure alone does not necessarily lead to poor LAVI performance. However, when fragmentation is dominated by a small number of specific inter-community connections, cumulative half-link loss may not adequately represent the structural consequences of their removal. The bio-diseasome network illustrates this limitation: nodes with similar cumulative half-link-removal profiles can yield markedly different fragmentation outcomes because the disruption of particular bridge-like connections can strongly affect global connectivity. Thus, LAVI should be interpreted as a proxy for structural vulnerability rather than as a direct measure of fragmentation.

These findings highlight that the effectiveness of LAVI is inherently topology-dependent, reflecting the extent to which early cumulative connectivity loss serves as a reliable proxy for global fragmentation under localized attack dynamics. A quantitative topological threshold for the reliability of LAVI is not derived in the present study. Existing analytical frameworks for localized attacks typically characterize shell structure and percolation properties from a randomly selected root in random networks with a prescribed degree distribution~\cite{shao2015percolation,dong2019localized}. In contrast, the vulnerability considered here is conditioned on a specific seed node and therefore depends on its particular shell structure and on structural correlations and meso-scale organization that cannot generally be determined from the degree distribution alone. Developing a seed-conditioned analytical framework that incorporates structural correlations and meso-scale organization remains an important direction for future work. Relatedly, an effective attack radius may be introduced by defining a distance-truncated version of LAVI, $\mathcal{L}^{(d)}(i)$, that incorporates only shells within a maximum distance $d$ from the seed. Such an extension may preserve seed-specific information in the local neighborhood while reducing the contribution of distant shells that increasingly reflect global network structure. Determining an appropriate cutoff $d$, potentially in relation to characteristic path length or shell-growth properties, is left for future investigation.

A related limitation concerns the definition of locality. The localized attack considered in this study is defined by shortest-path distance on the network and therefore represents topological locality rather than geographical or physical distance. The additional RGG experiments examine a spatially embedded network topology, and the superior performance of closeness centrality in this case suggests that the global position of a seed may become particularly important when connectivity is spatially constrained. Nevertheless, the attack process itself remains defined in terms of topological distance. Extending the framework to localized attacks based explicitly on geographical distance or other physical constraints remains an important direction for future work.
 
It is important to emphasize that the present study focuses on vulnerability from a structural connectivity perspective and does not explicitly incorporate operational processes such as flow redistribution, capacity saturation, or application-specific performance measures. Although connectivity-based robustness captures essential structural degradation under localized attacks, real-world network functionality often depends on additional dynamic and weighted interaction mechanisms. Extending the proposed framework to incorporate such effects therefore represents an important direction for future research.

From a computational perspective, computing LAVI for a single seed node is relatively efficient, as it mainly relies on a breadth-first search and simple degree-based accumulation. However, applying LAVI to dynamically evolving networks poses additional challenges. In real-time settings where network topology changes frequently, repeated recomputation would be required, which may limit practical deployment. Moreover, the proposed framework assumes access to global network structural information, which may be difficult to obtain in systems where connectivity data are incomplete or distributed. At present, LAVI is therefore most suitable for static or slowly evolving networks, such as pre-disaster infrastructure models, highlighting the need for future extensions toward dynamic and partially observable network settings.

From an applied perspective, LAVI may be useful for identifying locations that are likely to experience severe damage under localized disruptions, such as regions highly exposed to earthquake-induced failures in transportation or utility networks, as well as structurally central regions in criminal or covert networks. By detecting seed nodes whose surrounding regions are particularly vulnerable, the proposed metric can support preventive protection strategies and targeted reinforcement of critical areas. For example, focusing on the localized neighborhoods surrounding nodes with high LAVI values may provide structural guidance for designing interventions to disrupt criminal or covert networks~\cite{ficara2023human}, while in infrastructure systems it may help prioritize regions for reinforcement or monitoring to mitigate the impact of localized hazards~\cite{zhou2019connectivity}.

Overall, our findings highlight the need to shift vulnerability analysis toward metrics that explicitly incorporate the topological and structural progression of failure. By linking node importance to the actual dynamics of damage progression, LAVI offers a practical alternative to traditional centrality-based assessments.

\section*{Conclusion}

This study introduced the Localized Attack Vulnerability Index (LAVI), a node-level metric for assessing the impact of LAs on complex networks. In contrast to conventional centrality measures, LAVI captures the spatial and structural trajectory of damage as it spreads from a seed node.

Simulations on both synthetic and real-world networks show that LAVI is strongly associated with robustness loss under localized disruptions, consistently outperforming classical centrality metrics. These findings highlight the need for vulnerability measures that capture the propagation and accumulation of localized damage, an aspect overlooked by conventional centrality-based approaches.

The findings demonstrate that the position of the seed node, the origin of spatial damage, plays a critical role in determining network robustness. By quantifying this effect, LAVI provides a foundation for assessing the vulnerability of infrastructure systems and informing countermeasures against localized disasters, such as earthquakes and tsunamis.

\section*{Code Availability}
The minimal data set and accompanying code underlying the findings of this study are publicly available at Zenodo via \url{https://doi.org/10.5281/zenodo.21884803}.

\section*{Funding}
This study was conducted while M.C. was employed as a Project Assistant Professor in a Social Cooperation Research Program jointly established by The University of Tokyo and TechnoPro, Inc. No specific grant number was associated with this support. TechnoPro, Inc. (\url{https://www.technopro.com/}) provided financial support for the program. The sponsor had no role in the study design, data collection and analysis, decision to publish, or preparation of the manuscript.

\section*{Competing interests}
The authors have declared that no competing interests exist.

\end{document}